\documentclass[11pt]{article}
\usepackage{amssymb}
\usepackage{amsfonts}
\usepackage{amsmath}
\usepackage{graphicx}
\providecommand{\U}[1]{\protect\rule{.1in}{.1in}}
\newtheorem{theorem}{Theorem}

\usepackage{mathtools}
\usepackage{color}

\newcommand{\cutset}{\partial\mathcal{V}}
\newcommand{\cutsize}{\vert\partial\mathcal{V}\vert}
\newcommand{\cutsizeplus}
{\vert\partial^{+}\mathcal{V}\vert}

\usepackage[normalem]{ulem} %used to strike through text

\begin{document}
%%%%%%%%%%%%%%%%%%%%%%%%%%%%%%%%%%%
\title{The Simplex Geometry of Graphs}
\author{Karel Devriendt\thanks{Currently at \emph{the Mathematical Institute, University of Oxford, Oxford} and \emph{The Alan Turing Institute, London}. \newline\indent\indent\emph{e-mail}:
devriendt@maths.ox.ac.uk}~~and~Piet Van Mieghem\thanks{\emph{e-mail}:
p.f.a.vanmieghem@tudelft.nl}
\\
\small{\emph{Faculty of Electrical Engineering, Mathematics and Computer Science,}} \\ \small{\emph{Delft University of Technology, Delft, The Netherlands}}}
\date{July 17, 2018}
\maketitle
%%%%%%%%%%%%%%%%%%%%%%%%%%%%%%%%%%%

%%%%%%%%%%%%%%%%%%%%%%%%%%%%%%%%%%%
\begin{abstract}
Graphs are a central object of study in various scientific fields, such as discrete mathematics, theoretical computer science and network science. These graphs are typically studied using combinatorial, algebraic or probabilistic methods, each of which highlights the properties of graphs in a unique way. Here, we discuss a novel approach to study graphs: the simplex geometry (a simplex is a generalized triangle). This perspective, proposed by Miroslav Fiedler, introduces techniques from (simplex) geometry into the field of graph theory and conversely, via an \emph{exact correspondence}. We introduce this graph-simplex correspondence, identify a number of basic connections between graph characteristics and simplex properties, and suggest some applications as example.
\end{abstract}
\section{Introduction}
In this article, we review and further develop the work of Miroslav Fiedler \cite{Fiedler_book} on the connection between \emph{graphs} and \emph{simplices} (higher-dimensional triangles). In contrast to other concepts and techniques introduced by Fiedler, which are now a central part of (spectral) graph theory and network science, e.g. \cite{Fiedler_algebraic},\cite{Fiedler_eigenvector}, his work on simplex geometry and its connection to graphs seems to have gone largely unnoticed in these fields.
\\
In the introduction of his 2011 book `\emph{Matrices and graphs in geometry}', Fiedler \cite{Fiedler_book} states that simplex geometry, which was the subject of his 1954 thesis \cite{Fiedler_thesis1, Fiedler_thesis2, Fiedler_thesis3}, fascinated him ever since his student days. This lifelong interest led to an impressive body of work on simplex geometry and its relation to matrix theory and graph theory, two other celebrated expertises of Fiedler. His book \cite{Fiedler_book} summarizes these contributions and includes previously unpublished results. 
The particular subject we discuss in this article is an \emph{exact} geometric representation of graphs as simplices, where graph properties such as degrees, cuts, eigenvalues, etc. appear as geometric invariants of a simplex. As the results on this graph-simplex correspondence are spread out over Fiedler's book \cite{Fiedler_book} and his many papers on the subject, we hope that by collecting and reviewing them in this article, we can give a more focused and structured overview of the topic. Since we have chosen to give a self-contained description of Fiedler's results, the breadth of this article is unfortunately limited to describing the correspondence and a number of basic results. This should, however, enable the reader to understand the basic principles of the graph-simplex correspondence, and serve as an introduction and supplement to the reading of \cite{Fiedler_book}. It is our hope that this exposition of Fiedler's geometric approach to graph theory, may convince the reader of its promising potential, and stimulate further research in this direction. 
\\
\\
Apart from Fiedler's work, there exist numerous other approaches to study graphs in a metric or geometric setting. We are not able to provide a full overview here, but will discuss a small selection of the existing alternative approaches.
\\
The best known and probably most natural distance function on a graph, is the shortest-path distance. This distance function is widely studied in graph theory \cite{Harary}, and typical and extremal distances are well understood in many classes of graphs. Moreover, the observation of remarkably small distances between nodes in many real-world networks \cite{Watts} was one of the landmark results that started the development of a whole new field of research, now called network science \cite{Newman}. While a graph with the shortest-path distance is generally not embeddable in Euclidean space, approximate low-distortion embeddings in low dimensions are often used \cite{Bourgain, Linial} to study and solve algorithmic problems on graphs. 
\\
Another important distance function on graphs is the effective resistance \cite{Klein}, also called resistance distance. Originally a concept in electrical circuit theory, the effective resistance is intimately related to random walks on graphs \cite{Nash-Williams, Doyle} and was shown to determine a metric, or distance function on graphs \cite{Klein}. While a graph with the effective resistance as distance function is generally not embeddable in Euclidean space, the square root of the effective resistance is equal to the Euclidean distance \cite{Klein_wiener}. In Section \ref{S5} we briefly discuss how the effective resistance appears naturally in relation to the graph-simplex correspondence.
\\
Lov\'asz \cite{Lovasz} introduced the concept of orthogonal graph representations, where a vector in Euclidean space is assigned to each node in a graph, such that non-adjacent nodes in the graph correspond to orthogonal vectors. The graph-simplex correspondence described in this article fits the concept of an orthogonal graph representations, but to the best of our knowledge, simplex geometry and Fiedler's correspondence in particular have not been investigated in the context of orthogonal graph representations.
\\
A more recent development is the embedding \cite{Serrano} of real-world networks into 'hidden' geometric spaces, where nodes are assumed to be positioned in a geometric space and have their connections determined (probabilistically) by their proximity to other nodes in this space. Interestingly, certain geometries and in particular hyperbolic geometry, give rise to graph ensembles with typical real-world features such as small-worldness, clustering and broad degree distributions \cite{Krioukov}. The difference between the simplex approach and the (hyperbolic) embedding of real-world networks is that the latter is a low-dimensional approximation for graphs, which captures the main features of a real-world network in the geometric properties of the underlying space, while the graph-simplex correspondence exactly translates a graph's structure into a high dimensional, though simple geometric object.
\\
As with every new perspective, we expect that the graph-simplex correspondence will lead to interesting new questions and insights in the properties of graphs and, hopefully, may contribute to the resolution of open challenges and problems.
\\
\\
In Section \ref{S2}, the two fundamental objects of interest are introduced: graphs and simplices. Next, in Section \ref{S3}, Fiedler's graph-simplex correspondence is described. In Section \ref{S4}, a number of graph properties and their correspondence in the simplex geometry are discussed: degree, generalized degree (cut size), Laplacian eigenvalues and finally the number of spanning trees. In Section \ref{S5}, we conclude the article and summarize the results. A list of symbols can be found in Appendix \ref{Alist}.
%%%%%%%%%%%%%%%%%%%%%%%%%%%%%%%%%%%

%%%%%%%%%%%%%%%%%%%%%%%%%%%%%%%%%%%
\section{Preliminaries}\label{S2}
\subsection{Graphs and the Laplacian matrix}
A graph $G(\mathcal{N},\mathcal{L})$ consists of a set $\mathcal{N}$ of $N$ nodes and a set $\mathcal{L}$ of $L$ links that connect pairs of distinct nodes. A common way to represent undirected graphs is by the $N\times N$ Laplacian matrix $Q$ with elements
$$
(Q)_{ij} = 
\begin{cases}
d_i &\text{~if~} i=j \\
-1 &\text{~if~} (i,j)\in\mathcal{L} \\
0 &\text{~otherwise},
\end{cases}
$$
where the degree $d_i$ is equal to the number of nodes adjacent to node $i$. In the case of weighted graphs, each link $(i,j)\in\mathcal{L}$ also has an associated weight $w_{ij}>0$ and the degree is equal to the sum of all incident link weights. An unweighted graph is thus a special case of a weighted graph with all link weights equal to $w_{ij}=1$. The pseudoinverse $Q^{\dagger}$ of the Laplacian matrix $Q$ is defined by the relations \cite{krl_bestspreader}
$$
QQ^{\dagger}=Q^{\dagger}Q = I-\frac{uu^T}{N},
$$
where $u=(1,1,\dots,1)^T$ is the all-one vector. As suggested by its name, the \emph{pseudoinverse Laplacian} $Q^{\dagger}$ is the inverse of the Laplacian matrix in the space orthogonal to $u$. In other words, the expression $Qx=y$ can be inverted to $Q^{\dagger}y=x$ when $u^Tx=u^Ty=0$ holds. Since many results for the Laplacian matrix $Q$ hold analogously for the pseudoinverse matrix $Q^{\dagger}$, we will use the superscript ``$+$'' to denote variables related to the pseudoinverse. For instance, the degree $d_i = (Q)_{ii}$ has the related pseudoinverse variable $d^{+}_i = (Q^{\dagger})_{ii}$. The superscript $(~)^+$ is thus part of the notation of a variable, while the superscript $(~)^\dagger$ denotes the pseudoinverse operator on a matrix.
\\
Since the Laplacian is a real and symmetric matrix, the solutions to the eigenvalue equation $Qz_k=\mu_kz_k$ are orthonormal eigenvectors $z_k$ and real eigenvalues $\mu_k$. The resulting eigendecomposition of $Q$ is then 
\begin{equation} \label{eq1}
Q = \sum_{k=1}^N \mu_k z_kz_k^T \quad\text{with~} \mu_k\in\mathbb{R}\text{~and~}z_k^Tz_m = \delta_{km},
\end{equation}
where $\delta_{km}$ is the Kronecker delta which is equal to $\delta_{km}=1$ if $k=m$ and zero otherwise. Introducing the $N\times N$ eigenvector matrix $\widetilde{Z} = [z_1~z_2~\dots z_N]$ and the $N\times N$ diagonal eigenvalue matrix $\widetilde{M} = \operatorname{diag}(\mu_1,\mu_2,\dots,\mu_{N})$, the eigendecomposition is compactly written as $Q=\widetilde{Z}\widetilde{M}\widetilde{Z}^T$.
\\
A fundamental result from spectral graph theory is that the Laplacian $Q$ of a connected undirected graph is positive semidefinite with a single eigenvalue equal to zero, and with the zero-eigenvector in the direction of the all-one vector \cite[art. 80]{pvm_GS}. By this result, we can denote the eigenvalues as an ordered set $\mu_1\geq\mu_2\geq\dots\geq\mu_{N-1}>\mu_N=0$, and the eigenvector $z_N=\frac{u}{\sqrt{N}}$. Furthermore, using the $N\times (N-1)$ matrix $Z = [z_1~z_2\dots~z_{N-1}]$ and the $(N-1)\times (N-1)$ matrix $M=\operatorname{diag}(\mu_{1},\mu_2,\dots,\mu_{N-1})$ with the zero eigenvalue $\mu_N$ and corresponding eigenvector $z_N$ omitted, we can write the eigendecomposition of the Laplacian as
$$
Q = \sum_{k=1}^{N-1}\mu_k z_kz_k^T \quad\text{~and~}\quad Q=ZMZ^T.
$$
Similarly, the pseudoinverse Laplacian $Q^{\dagger}$ is also a symmetric positive semidefinite matrix \cite{krl_bestspreader}, and has the eigendecomposition:
$$
Q^{\dagger} = \sum_{k=1}^{N-1}\frac{1}{\mu_k}z_kz_k^T\quad\text{~and~}\quad Q^{\dagger}=ZM^{-1}Z^T.
$$
%-------------------------------------------
\subsection{The Simplex}\label{S2.2}
A \emph{simplex} $\mathcal{S}$ is a geometric object that generalizes triangles and tetrahedra to any dimension. In $D=0,1,2,3$ dimensions, a simplex corresponds to a point, a line segment, a triangle and a tetrahedron, as shown in Figure \ref{fig1} below. 
\begin{figure}[h!]
\centering
\includegraphics[scale = 0.77]{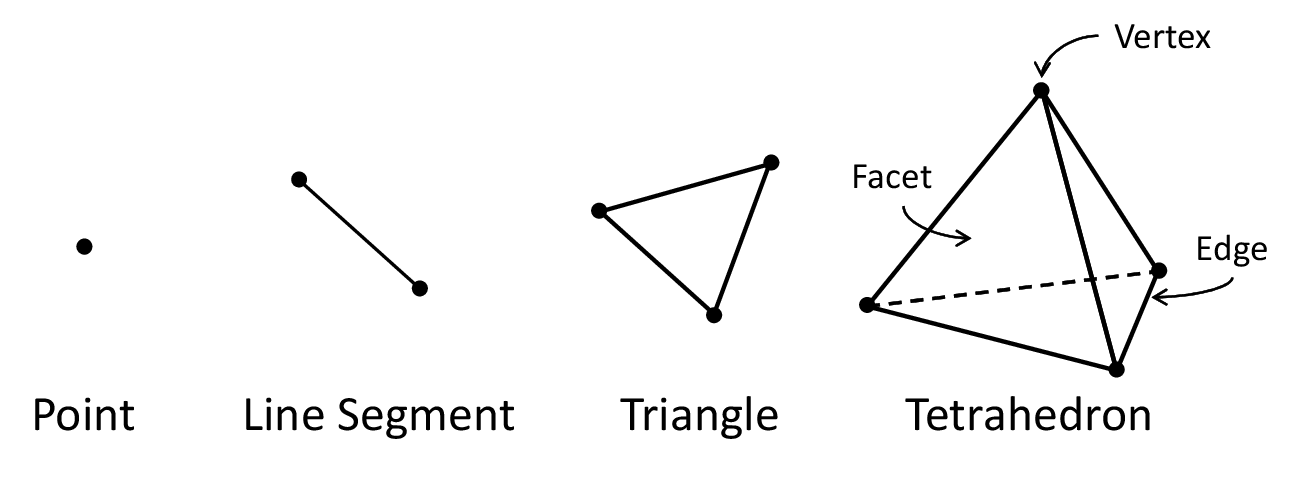}
\caption{Examples of low-dimensional simplices in $D=0,1,2,3$ dimensions.}
\label{fig1}
\end{figure}
Figure \ref{fig1} also illustrates that a simplex in $D$ dimensions is determined by $D+1$ points, which are called the \emph{vertices} of the simplex. The $i^\text{th}$ vertex is denoted by the vector $s_i\in\mathbb{R}^D$, and all vertices of a simplex are compactly represented by the $D\times(D+1)$ \emph{vertex matrix} $S=[s_1~s_2~\dots~s_{D+1}]$ containing the $D+1$ vertex vectors of $\mathcal{S}$ as columns. In order to determine a simplex, the vertex vectors $s_i$ need to satisfy certain independence relations similar to ``three \emph{non-collinear} points in a plane determine a triangle''. The independence condition states that $S$ must have rank $D$ in order for these $D+1$ vertices to determine a simplex. A more specific description of simplices, is that \emph{a simplex is the convex hull of its vertices}. This means that $\mathcal{S}$ is the set of all points $p\in\mathbb{R}^D$ that can be expressed as a \emph{convex combination} of its vertices $s_i$:
\begin{equation} \label{eq2}
\mathcal{S} = \left\lbrace p\in\mathbb{R}^{D} \bigg\vert p = Sx \textup{~with~} (x)_i\geq 0 \textup{~and~} u^Tx=1 \right\rbrace,
\end{equation}
where the vector $x\in\mathbb{R}^{D+1}$ determines the convex coefficients of $p$ with respect to the vertices $s_i$, i.e. all entries of the vector $x$ are non-negative and sum to one. The fact that any point in the simplex can be expressed as a linear combination of its vertices as $p=Sx$, is important in studying the simplex using algebraic methods, and the vector $x$ is called the \emph{barycentric coordinate} of the point $p$, with respect to the simplex $\mathcal{S}$. Figure \ref{fig2} exemplifies how barycentric coordinates specify the location of a point $p$ on the simplex, based on the vertex vectors $s_1,s_2$ and $s_3$.
\begin{figure}[h!]
\centering
\includegraphics[]{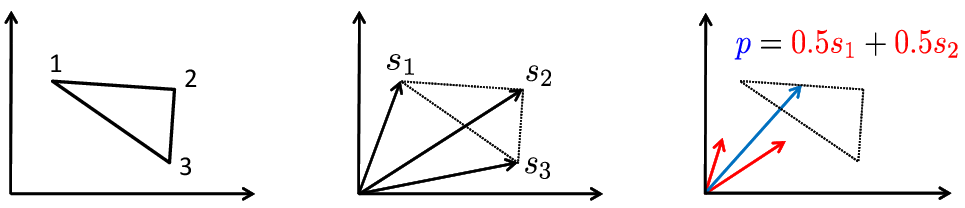}
\caption{All points on a simplex can be specified using barycentric coordinates, which determine its position as a convex combination of the simplex vertices.}
\label{fig2}
\end{figure}
\\
\\
As illustrated in Figure \ref{fig1}, the surface or boundary of a simplex consists of lower-dimensional simplices. In general, these basic constituents of the surface are called the \emph{faces} of the simplex, and each of these faces is also a $D_f$-dimensional simplex, with $0\leq D_f<D$. Some faces have a specific name: a $0$-dimensional face corresponds to a vertex, a $1$-dimensional face is commonly called an \emph{edge} and a $(D-1)$-dimensional face is called a \emph{facet}. Specifically, a $D_f$-dimensional face is the convex hull of $V\triangleq D_f+1$ vertices of the simplex, and if we denote the index-set that determines these vertices by $\mathcal{V}$, then the face $\mathcal{F}_\mathcal{V}$ determined by these vertices is defined as:
$$
\mathcal{F}_{\mathcal{V}} = \left\lbrace p\in\mathbb{R}^D \bigg\vert p=Sx_{\mathcal{V}} \textup{~with~} (x_{\mathcal{V}})_i \geq 0 \text{~and~} u^Tx_{\mathcal{V}}=1\right\rbrace,
$$ 
where the vector $x_{\mathcal{V}}\in\mathbb{R}^{D+1}$ denotes a barycentric coordinate with non-zero coefficients only for vertices in the set $\mathcal{V}$:
$$
\begin{cases} (x_{\mathcal{V}})_i\geq 0\quad\text{if~} i\in\mathcal{V} \\ (x_{\mathcal{V}})_i=0 \quad\text{if~} i\notin\mathcal{V}. \end{cases}
$$
Figure \ref{fig3} shows some faces of a tetrahedron. We further use $\mathcal{N}=\lbrace 1,2,\dots,D+1\rbrace$ to refer to the set of all vertex indices, $\mathcal{V}\subset\mathcal{N}$ to denote a subset of $V$ vertices, and $\bar{\mathcal{V}}=\mathcal{N}\backslash\mathcal{V}$  for the complementary set of vertices. A pair of faces that are determined by complementary vertex sets, e.g. $\mathcal{F}_{\mathcal{V}}$ and $\mathcal{F}_{\bar{\mathcal{V}}}$, are called \emph{complementary faces}.
\begin{figure}[h!]
\centering
\includegraphics[scale = 0.6]{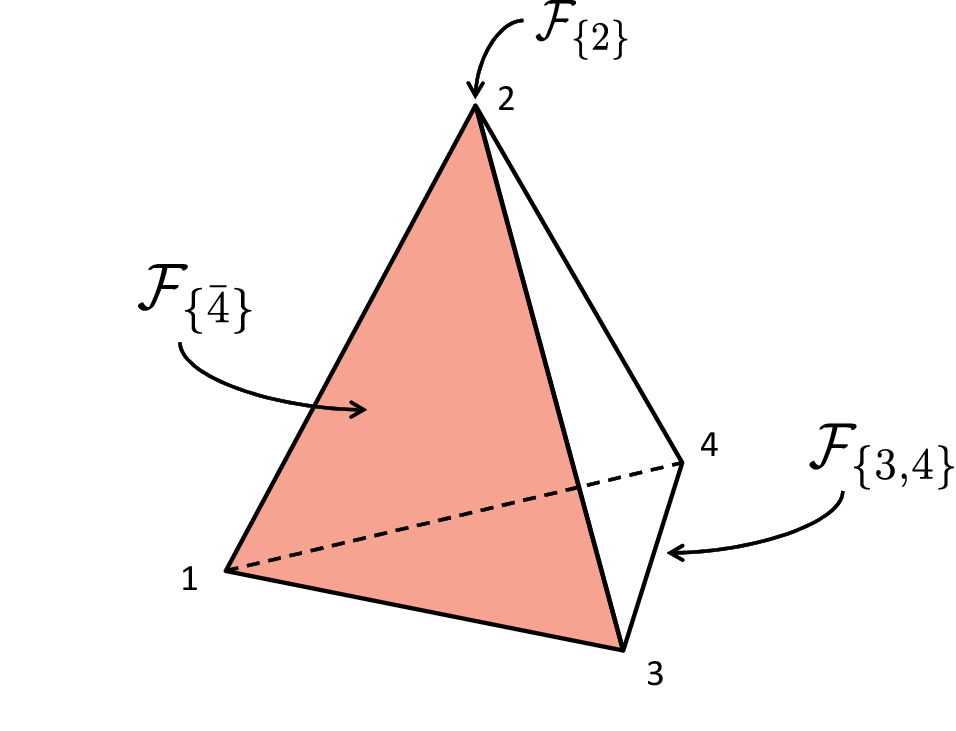}
\caption{Faces of a simplex}
\label{fig3}
\end{figure}
\\
\\
\textbf{To summarize:} a simplex $\mathcal{S}$ in $D$ dimensions is the convex hull of $D+1$ vertices $s_1,s_2,\dots,s_{D+1}$. The boundary of $\mathcal{S}$ consists of faces $\mathcal{F}_\mathcal{V}$, which are $D_f$-dimensional simplices determined by a subset $\mathcal{V}\subset\mathcal{N}$ containing $V=D_f+1$ vertices. The introduced symbols are summarized in Appendix \ref{Alist}.
%%%%%%%%%%%%%%%%%%%%%%%%%%%%%%%%%%%%%%%%%%%%%%%%%
\section{The Graph-Simplex correspondence}\label{S3}
In his 1976 paper `\emph{Aggregation in graphs}' \cite{Fiedler_aggregation}, Fiedler proved that every connected, undirected graph on $N$ nodes corresponds to one specific simplex $\mathcal{S}$ in $D=N-1$ dimensions\footnote{A similar statement is true in the reverse direction: every simplex with non-obtuse angles (smaller than or equal to $\tfrac{\pi}{2}$ radians) between all pairs of facets, is the inverse simplex of a connected, undirected graph with positive link weights (see Section \ref{S3a} for the definition of an inverse simplex).}. As Fiedler \cite{Fiedler_Steiner} points out, this graph-simplex correspondence means that ``\emph{every geometric invariant of the simplex is at the same time an invariant of the graph}''. Much of his later work in simplex geometry focuses on this pursuit of connecting simplex properties to graph properties. Here, we will show how the correspondence between a graph $G$ on $N$ nodes and a simplex $\mathcal{S}$ in $N-1$ dimensions can be studied explicitly using the Laplacian matrix $Q$. The connection between a graph and a simplex \cite{Fiedler_Steiner} is then immediate: \emph{The Laplacian matrix $Q$ of a graph is the Gram matrix of the vertex vectors $s_i$ of a simplex $\mathcal{S}$}. A Gram matrix of a set of vectors $p_i\in\mathbb{R}^N$ is the positive semidefinite matrix $X$ with elements equal to the inner product between pairs of points, i.e. $(X)_{ij} = p_i^Tp_j$. Hence, the vertex vectors $s_i$ of the simplex $\mathcal{S}$ are related to the Laplacian matrix $Q$ by
\begin{equation} \label{eq3}
(Q)_{ij} = s_i^Ts_j \quad\text{or}\quad Q = S^TS,
\end{equation}
which uniquely defines the $N$ vertices $s_i\in\mathbb{R}^{N-1}$, the $N\times(N-1)$ vertex matrix $S$, and thus the simplex $\mathcal{S}$. A more explicit expression for the vertex vectors $s_i$ follows from the eigendecomposition \eqref{eq1} of the Laplacian matrix: $Q = ZMZ^T = (Z\sqrt{M})(Z\sqrt{M})^T.$ Combined with \eqref{eq3}, the eigendecomposition \eqref{eq1} of the Laplacian thus specifies the vertex vectors $s_i$ as:
\begin{equation} \label{eq4}
S = (Z\sqrt{M})^T \quad\text{or}\quad (s_i)_k = (z_k)_i\sqrt{\mu_k}.
\end{equation}
Since every Laplacian matrix $Q$ allows the eigendecomposition \eqref{eq1}, expression \eqref{eq4} indeed assigns a unique set of $N$ vertices $s_i$ to each graph. However, it is not obvious that these vertices actually determine a simplex. This specific property of the vertices follows from the fact that rank($S$) = rank($S^TS$) = rank($Q$) = $N-1$, which means that the vertices $s_i$ are independent (in the sense introduced in Section \ref{S2.2}) and thus determine a simplex. Figure \ref{fig4} below illustrates the graph-simplex correspondence for an example with $N=4$ nodes. As an additional numerical example, we consider the path graph on four nodes $P_4$ (the leftmost graph in Figure \ref{fig4b}). The Laplacian matrix $Q$, eigenvector matrix $Z$ and eigenvalue matrix $M$ (with the constant eigenvector and zero eigenvalue omitted, respectively) of the path graph are equal to:
$$
Q = \begin{bmatrix}
1 & -1 & 0 & 0 \\
-1 & 2 & -1 & 0 \\
0 & -1 & 2 & -1 \\
0 & 0 & -1 & 1
\end{bmatrix}
\text{~with~}
Z = \begin{bmatrix}
0.653   & 0.5  & -0.271 \\
0.271   & -0.5 & 0.653 \\
-0.271  & -0.5 & -0.653 \\
-0.653  & 0.5  & 0.271 
\end{bmatrix}
\text{~and~}
M = \begin{bmatrix}
0.586 & 0 & 0 \\
0 & 2 & 0 \\
0 & 0 & 3.414
\end{bmatrix},
$$
with values rounded to three decimal precision. The vertex matrix $S$ of the simplex corresponding to the path graph $P_4$ is directly calculated from these matrices as $S = \sqrt{M}Z^T$:
$$
S = \begin{bmatrix}
0.5 & 0.207 & -0.207 & -0.5 \\
0.707 & -0.707 & -0.707 & 0.707 \\
-0.5 & 1.207 & -1.207 & 0.5
\end{bmatrix}.
$$
\begin{figure}[h!]
\centering
\includegraphics[scale = 0.8]{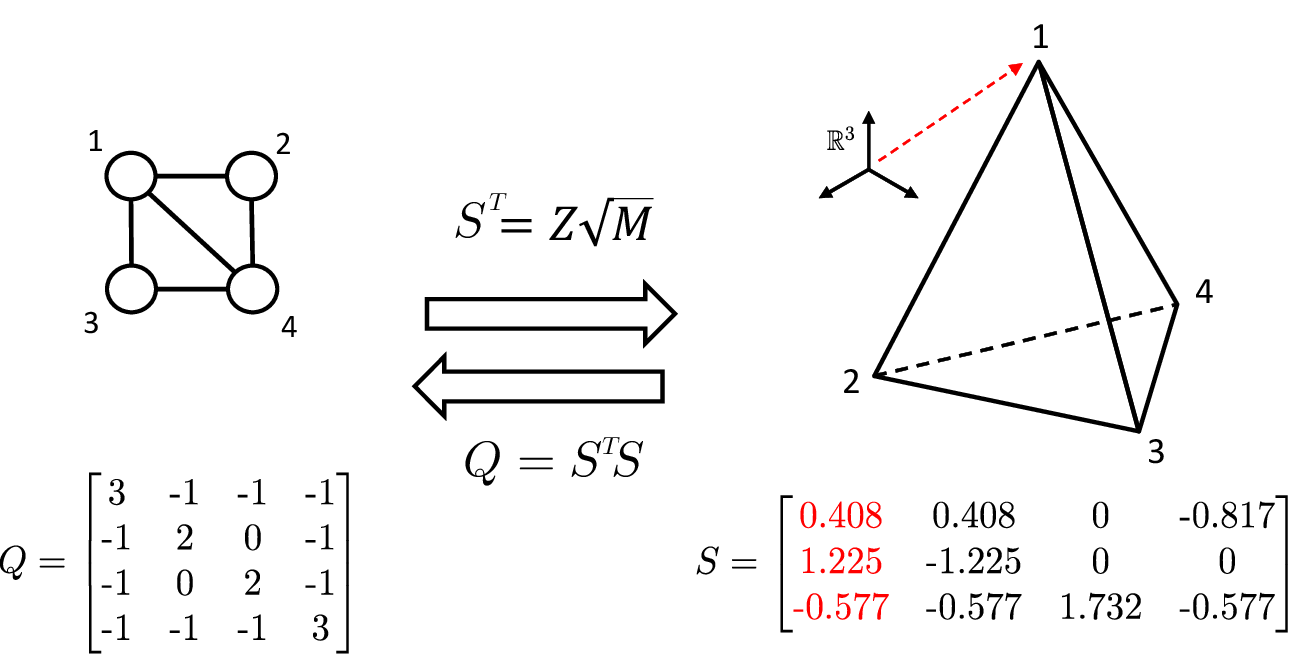}
\caption{Schematic overview of the graph-simplex correspondence.}
\label{fig4}
\end{figure}
Figure \ref{fig4b} shows all connected, unweighted graphs on four nodes and their corresponding tetrahedra. The collection of graph-simplex pairs in Figure \ref{fig4b} also highlights how similarity between nodes, i.e. sets of nodes which are indistinguishable with respect to their connection to the rest of the graph, is reflected in similarity between vertices in the simplex. This similarity is exemplified by nodes $\lbrace 2,3\rbrace$ in the path graph $P_4$, and the corresponding vertices $s_2,s_3$ in the tetrahedron (and likewise for nodes $\lbrace{1,4}\rbrace$). The most extreme example of similarity between nodes is achieved in the complete graph $K_4$, where all nodes are indistinguishable. Consequently, the vertices of the simplex $\mathcal{S}_{K_4}$ corresponding to the complete graph are also indistinguishable, which means that $\mathcal{S}_{K_4}$ and more generally $\mathcal{S}_{K_N}$ for any $N$ is a regular simplex; in other words, all edge-lengths and angles between facets of $\mathcal{S}_{K_N}$ are the same.
\begin{figure}[h!]
    \centering
    \includegraphics[width = \linewidth]{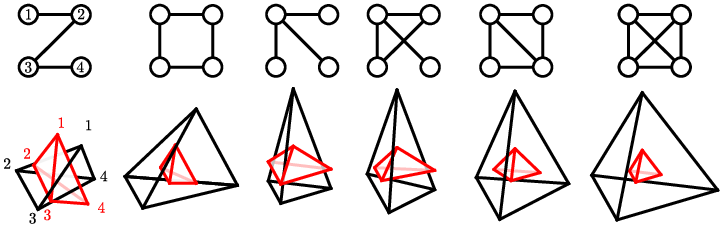}
    \caption{All connected, unweighted graphs on four nodes and their corresponding simplices (in black) and inverse simplices (in red, see Section \ref{S3a}). All simplices and inverse simplices are drawn on the same scale, and nodes and vertices are labelled as in the leftmost graph-simplex pair.}
    \label{fig4b}
\end{figure}
%------------------------------------------------------------------
\subsection{The inverse simplex of a graph} \label{S3a}
\begin{figure}[h!]
\centering
\includegraphics[scale = 0.68]{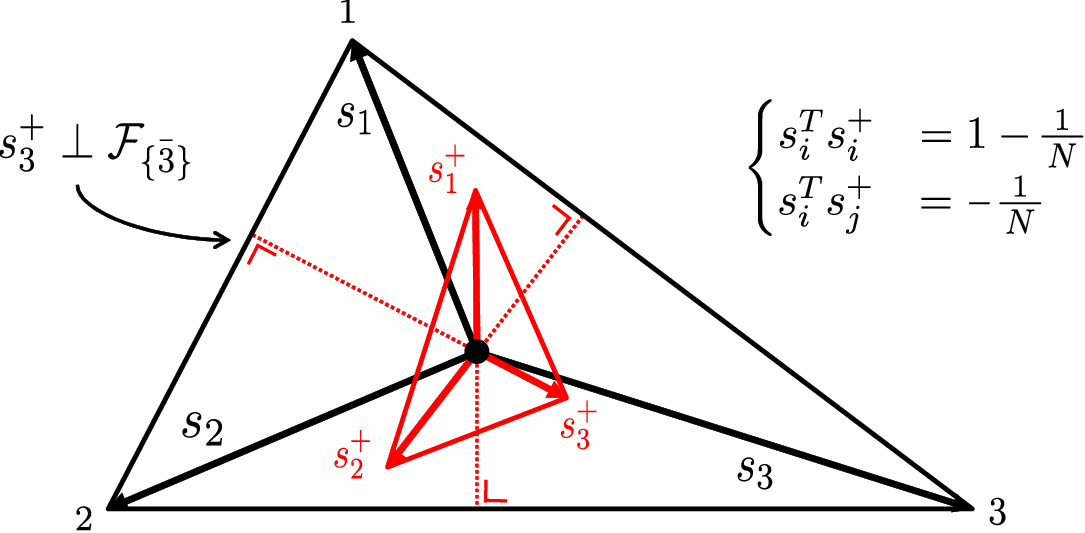}
\caption{The vertex vectors of a simplex $\mathcal{S}$ (in black) are parallel to the inner normal vectors of the facets of its inverse simplex $\mathcal{S}^{+}$ (in red). The black dot represents the origin of $\mathbb{R}^2$, which coincides with the centroid of both triangles (see Section \ref{S4.1}).}
\label{fig5}
\end{figure}

Fiedler \cite{Fiedler_involution} introduced the concept of an \emph{inverse simplex} of a graph, based on the (bi)orthogonal relations between a matrix and its pseudoinverse (see also \cite[Chapter 5.1]{Fiedler_book}). The inverse simplex $\mathcal{S}^{+}$ of a graph $G$ is defined as the simplex whose vertices $s^{+}_i$ have the pseudoinverse Laplacian $Q^{\dagger}$ as Gram matrix. In other words, the inverse simplex $\mathcal{S}^{+}$ is the convex hull of the vertices $s^{+}_i$ defined by:
\begin{equation}\label{eq5}
S^{\dagger} = Z\sqrt{M^{-1}} \quad\text{or}\quad (s^{+}_i)_k = (z_k)_i\frac{1}{\sqrt{\mu_k}}.
\end{equation}
To illustrate the inverse simplex concept, Figure \ref{fig4b} shows the simplex and inverse simplex of all four-node graphs and Figure \ref{fig5} shows a pair of inverse triangles and their vertex vectors in more detail. In order to clearly distinguish between these two simplices related to a graph $G$, we will further refer to $\mathcal{S}$ as the \emph{original simplex} and to $\mathcal{S}^{+}$ as the \emph{inverse simplex} of $G$.
From the definition \eqref{eq4} of the simplex vertices $s_i$ and the inverse simplex vertices $s^{+}_i$ in \eqref{eq5}, we find that their inner products satisfy
\begin{equation} \label{eq5b}
s_i^Ts^{+}_j = 
\begin{cases}
1-\tfrac{1}{N} &\text{~if~} i=j \\
-\tfrac{1}{N} &\text{~otherwise}.
\end{cases}
\end{equation}
As a result of \eqref{eq5b}, the vertex matrices $S$ and $S^{\dagger}$ satisfy the pseudoinverse relations:
\begin{equation} \label{eq6}
S^{\dagger T}S = S^TS^{\dagger} = I - \frac{uu^T}{N}.
\end{equation}
From these pseudoinverse relations \eqref{eq6}, the interesting result follows that the vertex vector $s^{+}_i$ of the inverse simplex $\mathcal{S}^{+}$ is parallel to the inner normal vector of the facet $\mathcal{F}_{\bar{\lbrace i\rbrace}}$ (see also Figure \ref{fig5}). In other words, 
$s^{+}_i$ is orthogonal to any vector that points from one point in $\mathcal{F}_{\bar{\lbrace i\rbrace}}$  to another:
\begin{equation} \label{eq7}
s_i^{+ T}(p-q) = 0, \text{~for all~} p,q\in\mathcal{F}_{\bar{\lbrace i \rbrace}}.
\end{equation}
Similarly, the vertex vector $s_i$ of the original simplex $\mathcal{S}$ is parallel to the inner normal vector of the facet $\mathcal{F}^{+}_{\bar{\lbrace i\rbrace}}$ of the inverse simplex $\mathcal{S}^{+}$. Expression \eqref{eq7} can be checked by using the barycentric coordinates of the points in $\mathcal{F}_{\bar{\lbrace i \rbrace}}$ as $p = Sx_{\bar{\lbrace i\rbrace}}$ and $q =Sy_{\bar{\lbrace i\rbrace}}$, and invoking the pseudoinverse relation \eqref{eq6} between $S$ and $S^{\dagger}$, which gives $s^{+ T}_iSx_{\bar{\lbrace i\rbrace}} = s^{+ T}_iSy_{\bar{\lbrace i\rbrace}} = -\tfrac{1}{N}$.
\\
Another interesting consequence of the pseudoinverse relation \eqref{eq6} between the vertex matrices $S$ and $S^{+}$, and the fact that the vertex vectors of the inverse simplex $\mathcal{S}^{+}$ determine the normal direction of the facets of the original simplex $\mathcal{S}$, is that it enables a compact dual definition of $\mathcal{S}$. Each convex polytope $\mathcal{P}$ in $N-1$ dimensions has two dual definitions: either as the convex hull of a set of points $p_i\in\mathbb{R}^{N-1}$, or as the intersection of a number of halfspaces $\lbrace p\in\mathbb{R}^{N-1} \mid p^Tx \geq \alpha \rbrace_i$. In the case of a simplex\footnote{This dual definition holds true for general simplices, irrespective of any corresponding graph. For notational consistency we consider a simplex $\mathcal{S}$ in $N-1$ dimensions.}, definition \eqref{eq2} corresponds to the `convex hull' definition of $\mathcal{S}$, while using the vectors $s^{+}_i$ as facet normals, the `halfspace' definition of $\mathcal{S}$ follows as (see also Appendix \ref{A Halfspace}):
\begin{equation}\label{eq8}
\mathcal{S} = \left\lbrace p\in\mathbb{R}^{N-1} \bigg\vert p^TS^{\dagger} \geq -\frac{u}{N} \right\rbrace.
\end{equation}
These dual definitions highlight different aspects of the simplex. Definition \eqref{eq2}, for instance, shows how a point $p$ in the simplex $\mathcal{S}$ can be generated by using a barycentric coordinate as $p=Sx$. Definition \eqref{eq8} on the other hand, shows how to test whether a given point $p$ is inside the simplex, i.e. when $p^TS^{\dagger} \geq -\frac{u}{N}$ is satisfied.
\\
\\
\textbf{To summarize:} each graph $G$ corresponds to an original simplex $\mathcal{S}$ with vertex matrix $S$, whose Gram matrix is equal to the Laplacian $Q = S^TS$, and an inverse simplex $\mathcal{S}^{+}$ with vertex matrix $S^\dagger$, whose Gram matrix is equal to the pseudoinverse Laplacian $Q^{\dagger} = S^{\dagger T}S^\dagger$. The inverse simplices $\mathcal{S}$ and $\mathcal{S}^{+}$ satisfy orthogonal relations \eqref{eq7}, where the vertex vectors of one determine the inner normal directions of the other.
%----------------------------
\section{Related Graph and Simplex properties}\label{S4}
\subsection{Centroids of the simplex}\label{S4.1}
\begin{figure}[h!]
\centering
\includegraphics[scale = 0.50]{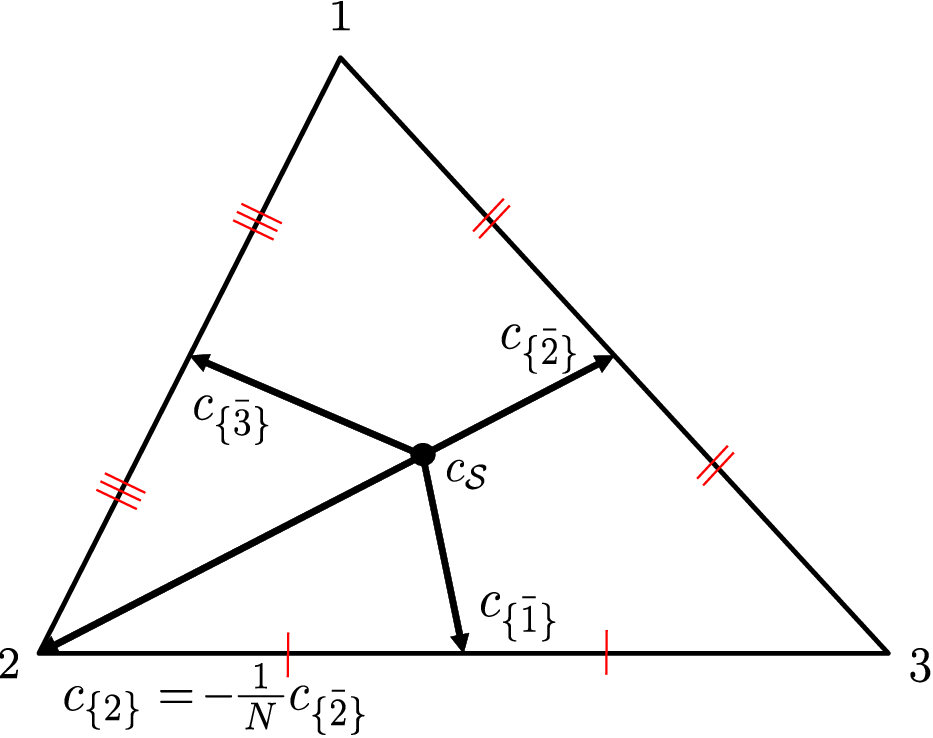}
\caption{Centroids in a simplex. The centroids of complementary faces are antiparallel, and satisfy equation \eqref{eq10}.}
\label{fig6}
\end{figure}

Before presenting the graph-simplex relations, we introduce a property of the center of mass of the simplex of a graph. The center of mass of a simplex, further called the \emph{centroid} of $\mathcal{S}$ and denoted by $c_\mathcal{S}$, is the arithmetic mean of all points that constitute the simplex. By linearity of the arithmetic mean and convexity of the simplex, we find that the centroid can be expressed using barycentric coordinates as $c_\mathcal{S} = S\frac{u}{N}$. In other words, the centroid of $\mathcal{S}$ is the (unique) point with its position determined by an equal convex combination of all vertices $s_i$. Since $S=\sqrt{M}Z^T$ and $Z^Tu=0$ hold for the simplex of a graph, we find the remarkable property that
$$
c_\mathcal{S} = S\tfrac{u}{N} = 0\in\mathbb{R}^{N-1} \colon \text{~The simplex centroid~}c_\mathcal{S}\text{~coincides with the origin of~}\mathbb{R}^{N-1}
$$
This means that all vectors in $\mathbb{R}^{N-1}$ have their ``starting point'' at $c_\mathcal{S}$. For instance, the vertex vectors $s_i$ of a simplex $\mathcal{S}$ are vectors pointing from the centroid $c_\mathcal{S}$ of $\mathcal{S}$ to the respective vertices. Moreover, definition \eqref{eq5} shows that $S^{\dagger} = \sqrt{M^{-1}}Z^T$, which means that the inverse-simplex centroid $c_{\mathcal{S}^+}$ also coincides with the origin $0\in\mathbb{R}^{N-1}$ and thus with the original simplex centroid $c_{\mathcal{S}}$. Indeed, in Figure \ref{fig5} the black dot indicates the centroid of both simplices $\mathcal{S}$ and $\mathcal{S}^{+}$.
\\
\\
Since each face of a simplex is a $(V-1)$-dimensional simplex, the vectors $c_{\mathcal{V}}\in\mathbb{R}^{N-1}$ pointing to the centroids of these faces also have a compact description in barycentric coordinates:
\begin{equation} \label{eq9}
c_\mathcal{V} = S\frac{u_\mathcal{V}}{V}\text{~is the centroid vector of the face~}\mathcal{F}_\mathcal{V}\text{, with~}
(u_\mathcal{V})_i=\begin{cases} 1\text{~if~}i\in\mathcal{V} \\ 0\text{~if~}j\notin\mathcal{V}\end{cases}
\end{equation}
Figure \ref{fig6} draws the centroids of a simplex and its faces. For complementary faces $\mathcal{F}_\mathcal{V}$ and $\mathcal{F}_{\bar{\mathcal{V}}}$, the centroid definition \eqref{eq9} together with the fact that $u_\mathcal{V} - u = u_{\bar{\mathcal{V}}}$, show that the centroid vectors $c_\mathcal{V}$ and $c_{\bar{\mathcal{V}}}$ are antiparallel and satisfy:
\begin{equation} \label{eq10}
-Vc_\mathcal{V} = (N-V)c_{\bar{\mathcal{V}}},
\end{equation}
The line between a pair of complementary centroids is also called a \emph{median}, and from \eqref{eq10} follows that such a median passes through the simplex centroid $c_\mathcal{S}$. Since this holds for every pair of complementary faces, the famous property follows that \emph{the medians of a simplex meet at its centroid}.
\\
Since there are ${N}\choose{V}$ faces of dimension $D_f=V-1$ and thus equally many centroids, there is a total of $\sum_{V=1}^{N-1} {{N}\choose{V}} = 2^N-2$ face centroids for a simplex in $N-1$ dimensions.
%----------------------------
\subsection{Local graph connectivity}
\begin{figure}[h!]
\centering
\includegraphics[scale = 0.6]{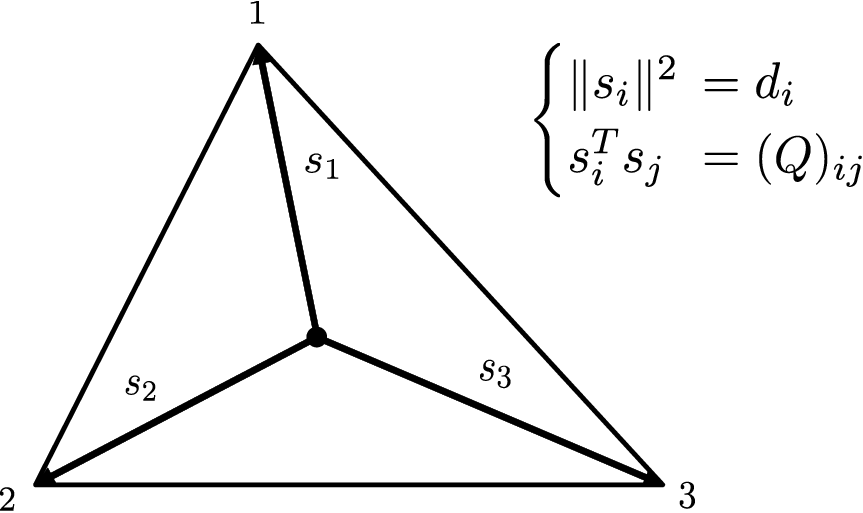}
\caption{The local connectivity structure of a graph, i.e. the degree and adjacency, can be deduced from the inner product between pairs of (possibly the same) vertex vectors.}
\label{fig7}
\end{figure}

The basic information that determines the local structure of a graph are the pairs of nodes $i,j$ that are connected by a link and, if applicable, the nonnegative weight $w_{ij}$ of this link. Given the simplex $\mathcal{S}$ of a graph $G$, the connectivity between a pair of (distinct) nodes $i$ and $j$, can be deduced from the inner product between the corresponding vertex vectors in the simplex:
\begin{equation} \label{eq11}
\begin{cases}
(i,j)\in\mathcal{L} &\text{~if~} s_i^Ts_j\neq 0 \\
(i,j)\notin\mathcal{L} &\text{~if~} s_i^Ts_j = 0 
\end{cases}
\quad\text{~for all~}i\neq j,
\end{equation}
which follows from the Gram relation \eqref{eq3}  between the Laplacian $Q$ and the vertex vectors, i.e. $s_i^Ts_j = (Q)_{ij}$. In the case of weighted graphs, the inner product of these vertex vectors is equal to the corresponding (negative) link weight $s_i^Ts_j = -w_{ij}$.
\\
A second local property of a graph is the degree of its nodes. Given the simplex $\mathcal{S}$ of a graph $G$, the degree of a node $i$ is related to the corresponding vertex vector $s_i$ as:
\begin{equation} \label{eq12}
\Vert s_i \Vert^2 = d_i,
\end{equation}
which again follows from \eqref{eq3}. In other words, the squared Euclidean distance from the simplex centroid $c_\mathcal{S}$ to one of its vertices $s_i$ is equal to the degree $d_i$ of the node corresponding to that vertex. Expression \eqref{eq11} and \eqref{eq12} hold analogously for the inverse simplex: $s_i^{+T}s^{+}_j = (Q^{\dagger})_{ij}$ and $\Vert s^{+}_i \Vert^2 = d^{+}_i$, where we introduce the notation $d^{+}_i = (Q^{\dagger})_{ii}$. Figure \ref{fig7} illustrates the two basic simplex-graph relations.
%-------------------------------------
\subsection{Global graph connectivity}
\begin{figure}[h!]
\centering
\includegraphics[scale = 1.2]{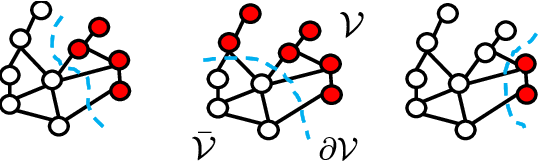}
\caption{Example of the cut set $\cutset$ for a number of different sets $\mathcal{V}$ in a graph.}
\label{fig8}
\end{figure}

\begin{figure}[h!]
\centering
\includegraphics[scale = 0.5]{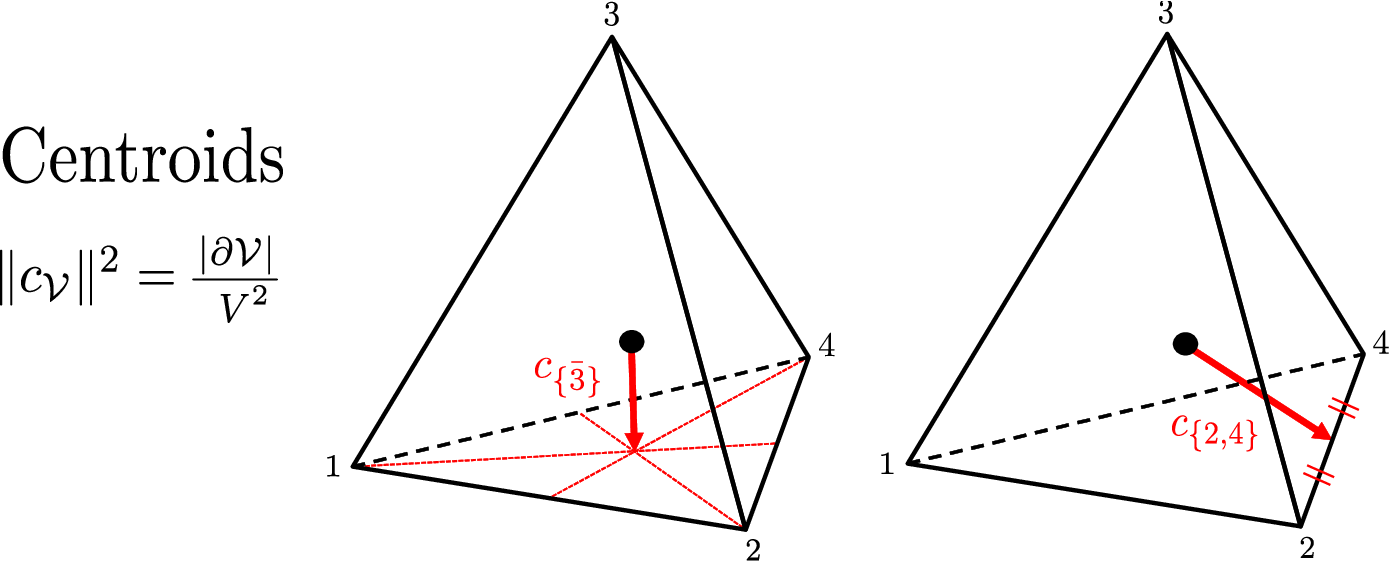}
\caption{Two centroid vectors in a simplex.}
\label{fig8b}
\end{figure}

\begin{figure}[h!]
\centering
\includegraphics[scale=0.5]{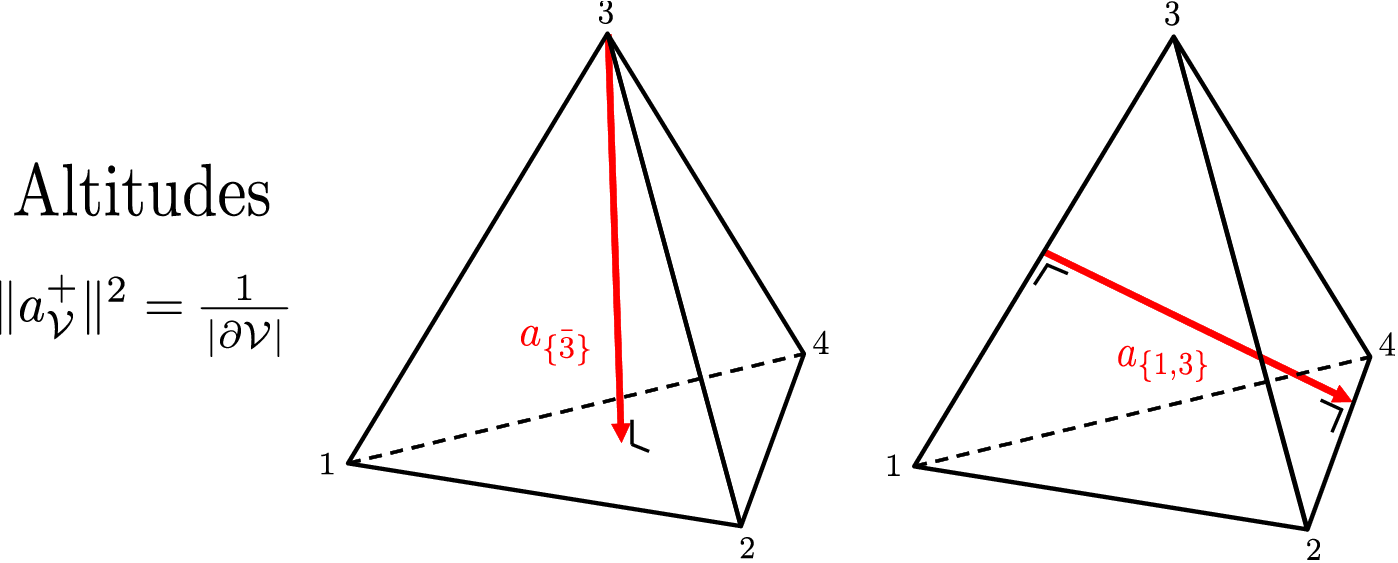}
\caption{Two altitudes in a tetrahedron.}
\label{fig9}
\end{figure}

As an extension of expressions \eqref{eq11} and \eqref{eq12} that identify local connectivity properties of a graph $G$ in the corresponding simplex $\mathcal{S}$, we show that global connectivity properties of $G$ are also identifiable in $\mathcal{S}$. Instead of the connectivity of nodes and pairs of nodes, we consider the connectivity of sets of nodes and between pairs of sets of nodes. If by $\mathcal{V}$ we denote a set of nodes in the graph, then the \emph{cut set} $\cutset$ (illustrated in Figure \ref{fig8}) is defined as the set of all links which connect nodes from $\mathcal{V}$ to nodes in $\bar{\mathcal{V}}$. In other words, the cut set $\cutset$ is defined as \cite{krl_report}:
$$
\cutset = \left\lbrace (i,j)\in\mathcal{L} \mid i\in\mathcal{V}\text{~and~}j\in\bar{\mathcal{V}} \right\rbrace.
$$
The number of links in a cut set is called the \emph{cut size}, and is denoted by $\cutsize$. The cut size of a set captures similar information as the degree of a node. In fact, the cut size reduces to the degree when $\mathcal{V}$ is a single node: $\vert\partial \lbrace i\rbrace\vert = d_i$.
The degree of a node $i$ is related by \eqref{eq12} to the length of the corresponding vertex vector $s_i$ of $\mathcal{S}$. Similarly, we find that the cut size $\cutsize$ of a set $\mathcal{V}$ is related to the (length of) the centroid vector $c_\mathcal{V}$ of the face $\mathcal{F}_\mathcal{V}$ as:
\begin{equation} \label{eq13}
\Vert c_\mathcal{V}\Vert^2 = \frac{\cutsize}{V^2} \quad\text{and}\quad \Vert c_{\bar{\mathcal{V}}} \Vert^2 = \frac{\vert\partial\bar{\mathcal{V}}\vert}{(N-V)^2},
\end{equation}
which reduces to \eqref{eq12} when $\mathcal{V}=\lbrace i\rbrace$. Expression \eqref{eq13} follows from the fact that the cut size $\cutsize$ can be expressed \cite{pvm_GS, krl_report} as a quadratic product of the Laplacian matrix $Q$:
$$
\cutsize = \sum_{(i,j)\in\mathcal{L}} ((u_\mathcal{V})_i-(u_\mathcal{V})_j)^2 = u_\mathcal{V}^TQu_\mathcal{V}.
$$ 
Since the centroids of $\mathcal{F}_\mathcal{V}$ and $\mathcal{F}_{\bar{\mathcal{V}}}$ have barycentric coordinates proportional to $u_\mathcal{V}$ and $u_{\bar{\mathcal{V}}}$, their length is proportional to the quadratic product $u_\mathcal{V}^TQu_\mathcal{V}$, from which \eqref{eq13} follows. The analogous results hold for the inverse simplex, i.e. $\Vert c^{+}_\mathcal{V}\Vert^2 = \tfrac{\vert\partial^{+}\mathcal{V}\vert}{V^2}$, where we introduce the notation $\vert\partial^{+}\mathcal{V}\vert = u_\mathcal{V}^TQ^{\dagger}u_\mathcal{V}$ in analogy with the cut size in the original simplex. Figure \ref{fig8b} shows two centroid vectors in a tetrahedron.
Another simplex property that relates to the cut size $\cutsize$ is the altitude. An \emph{altitude} of the simplex $\mathcal{S}$ is a vector which points from one face $\mathcal{F}_\mathcal{V}$ to the complementary face $\mathcal{F}_{\bar{\mathcal{V}}}$ and which is \emph{orthogonal} to both faces (see Figure \ref{fig9}). In other words, the altitude can be written as:
$$
a_\mathcal{V} = p^{\star} - q^{\star}, \text{~for some~}p^{\star}\in\mathcal{F}_{\bar{\mathcal{V}}},\text{~and~}q^{\star}\in\mathcal{F}_\mathcal{V},
$$
where $p^\star$ and $q^\star$ are such that $a_\mathcal{V}$ is orthogonal to both faces.
In Appendix \ref{A Altitude}, we show that the altitude $a_\mathcal{V}$ is parallel to the complementary centroid of the inverse simplex $c_{\bar{\mathcal{V}}}^{+}$, in other words that
\begin{equation} \label{eq13b}
\frac{a_\mathcal{V}}{\Vert a_\mathcal{V}\Vert} = \frac{c^{+}_{\bar{\mathcal{V}}}}{\Vert c^{+}_{\bar{\mathcal{V}}}\Vert}.
\end{equation}
Furthermore, we show in Appendix \ref{A Altitude} that \eqref{eq13b} leads to an explicit expression for the altitudes:
\begin{equation} \label{eq13c}
a_\mathcal{V} = \frac{N-V}{\cutsizeplus}c^{+}_{\bar{\mathcal{V}}} \quad\text{and}\quad a^{+}_\mathcal{V} = \frac{N-V}{\cutsize}c_{\bar{\mathcal{V}}},
\end{equation}
from which the length of the altitude $a^{+}_\mathcal{V}$ then follows as:
\begin{equation} \label{eq14}
\Vert a^{+}_\mathcal{V}\Vert^2 = \frac{1}{\cutsize}.
\end{equation}
Relation \eqref{eq14} generalizes the result of Fiedler \cite[Cor. 1.4.14]{Fiedler_book} that the altitude from a vertex $s^{+}_i$ to the complementary face $\mathcal{F}^{+}_{\bar{\lbrace i\rbrace}}$ in the inverse simplex has a length equal to the inverse degree of node $i$: $\Vert a^{+}_i\Vert^2 = \frac{1}{d_i}$. 
\\
Similar to how the cut set $\cutset$ of a set $\mathcal{V}$ generalizes the neighborhood of a node $i$, the intersection between two cut-sets $\partial\mathcal{V}_1\cap\partial\mathcal{V}_2$ can be seen as a generalization of the incidence between a pair of nodes $i$ and $j$:
$$
\partial\mathcal{V}_1 \cap \partial\mathcal{V}_2 = \left\lbrace (i,j)\in\mathcal{L} \mid i\in\mathcal{V}_1\text{~and~}j\in\mathcal{V}_2 \right\rbrace.
$$
The number of links in this set (or the sum of their weights) is the global analogue of the weight $w_{ij}$ of a single link, and can be deduced from the simplex geometry as the (scaled) inner product between the centroids $c_{\mathcal{V}_1}$ and $c_{\mathcal{V}_2}$, or from the inverse-simplex altitudes $a^{+}_{\mathcal{V}_1}$ and $a^{+}_{\mathcal{V}_2}$ as:
\begin{equation}\label{eq14b}
c_{\mathcal{V}_1}^Tc_{\mathcal{V}_2} = -\frac{\vert\partial\mathcal{V}_1\cap\partial\mathcal{V}_2\vert}{V_1V_2} \quad\text{~and~}\quad a^{+ T}_{\mathcal{V}_1}a^{+}_{\mathcal{V}_2} = -\frac{\left\vert\partial\mathcal{V}_1 \cap \partial\mathcal{V}_2\right\vert}{\vert\partial\mathcal{V}_1\vert\vert\partial\mathcal{V}_2\vert},
\end{equation}
Equation \eqref{eq14b} generalizes equation \eqref{eq11} for the local connectivity of a graph, which is also found from \eqref{eq14b} when $\mathcal{V}_1=\lbrace i\rbrace$ and $\mathcal{V}_2=\lbrace j\rbrace$.
\\
Expressions \eqref{eq13}-\eqref{eq14b} show the relation between graph-theoretic and geometric properties with a distinct combinatorial nature: the cut size $\cutsize$, face centroids $c_\mathcal{V}$ and altitudes $a_\mathcal{V}$, are all determined by one of the $2^N-2$ possible non-empty sets $\mathcal{V}\subset\mathcal{N}$. The relations between these properties has interesting implications. For the cut size, for instance, it is well known that finding the largest cut in a graph is NP-hard \cite{Karp}. Equation \eqref{eq14} then implies that finding extremal altitudes in a simplex suffers from the same problem of intractability. In particular, starting from NP-completeness of the Max-Cut problem \cite{Karp} and invoking equality \eqref{eq14}, we find:
\begin{gather} \label{eq14c}
\text{``Given~}G\text{~and~}k\in\mathbb{R}, \text{~is there a set~}\mathcal{V}\subset\mathcal{N}\text{~such that~}\cutsize \geq k \text{?''~ is NP-Complete}
\\
\Downarrow \nonumber
\\
\text{``Given~}\mathcal{S}^{+}\text{~and~}k\in\mathbb{R}, \text{~is there a set~}\mathcal{V}\subset\mathcal{N}\text{~such that~}\Vert a^{+}_\mathcal{V}\Vert \leq k \text{?''~ is NP-Complete}
\end{gather}
Importantly, $G$ should be a non-negatively weighted graph, and $S^{+}$ should be the inverse simplex of a non-negatively weighted graph $G$.
\\
\\
\textbf{To summarize:} the local connectivity of a graph $G$ -- the link weights $w_{ij}$ and degrees $d_i$ -- can be deduced from inner products of vector vectors $s_i$ of the simplex $\mathcal{S}$, following expression \eqref{eq11} and \eqref{eq12}. The global connectivity of a graph $G$ -- the size of cuts $\cutsize$ and cuts between a pair of sets $\vert\partial\mathcal{V}_1\cap\partial\mathcal{V}_2\vert$ -- can be deduced from (scaled) inner products of centroid vectors $c_{\mathcal{V}}$ and altitudes $a^{+}_\mathcal{V}$, following expression \eqref{eq13}, \eqref{eq14} and \eqref{eq14b}.
%------------------
\subsubsection{Geometric inequalities}
Since the altitude $a_\mathcal{V}$ between a pair of complementary faces $\mathcal{F}_\mathcal{V}$ and $\mathcal{F}_{\bar{\mathcal{V}}}$ is orthogonal to both faces, it is necessarily the shortest of all vectors lying between these faces. In other words, we obtain the inequality
\begin{equation} \label{eq15}
\Vert a_\mathcal{V} \Vert^2 \leq \Vert p-q\Vert^2, \text{~for all~} p\in\mathcal{F}_{\bar{\mathcal{V}}}\text{~and~}q\in\mathcal{F}_\mathcal{V}.
\end{equation}
If we translate this geometric inequality using barycentric coordinates, we obtain an inequality about quadratic products of the Laplacian $Q$ and its pseudoinverse $Q^\dagger$. The points $p$ and $q$ have barycentric coordinates $p=Sx_{\bar{\mathcal{V}}}$ and $q=Sx_\mathcal{V}$, such that the vector between them can be expressed as $p-q = S(x_{\bar{\mathcal{V}}}-x_\mathcal{V})=S\tilde{y}$ where $\tilde{y}$ is the `barycentric coordinate' of a vector pointing between complementary faces. More generally, any vector $y$ orthogonal to the all-one vector $u$ can be interpreted the (scaled) barycentric coordinate of a vector pointing between complementary faces:
$$
S\frac{y}{\Vert\tfrac{1}{2}y\Vert_1} = p - q, \text{~where~}p\in\mathcal{F}_{\mathcal{V}_y}\text{~and~}q\in\mathcal{F}_{\bar{\mathcal{V}}_y}\quad\forall y\in\mathbb{R}^N\text{~with~}u^Ty=0
$$
where $\mathcal{V}_y = \lbrace i\mid (y)_i\geq 0\rbrace$ is the set of non-negative entries of $y$, which determines in which faces the start-point and end-point of $S\tfrac{y}{\Vert \tfrac{1}{2}y\Vert_1}$ lie. Normalization by the $1$-norm $\Vert\tfrac{1}{2}y\Vert_1 = \tfrac{1}{2}\sum_{i=1}^N \vert (y)_i\vert$ is necessary to make the positive entries as well as the negative entries sum to one. Introducing this barycentric vector $y$ into the geometric inequality \eqref{eq15} of the altitude $a_\mathcal{V}$, we find:
\begin{theorem} \label{th1}
For any vector $y\in\mathbb{R}^N$ orthogonal to the all-one vector $u$ and with non-negative entries in the set $\mathcal{V}_y$, the quadratic product of $y$ with the Laplacian matrix $Q$ is bounded by:
\begin{equation}\label{eq15b}
y^TQy \geq \frac{\Vert\tfrac{1}{2}y\Vert_1^2}{\vert\partial^{+}\mathcal{V}_y\vert}
\end{equation}
\end{theorem}
In Appendix \ref{A 1-norm inequality} we provide an alternative derivation of Theorem \ref{th1} based on the Cauchy-Schwarz inequality invoked on the inner product $u_\mathcal{V}^Ty$.
\\
\\
As a Corollary of Theorem \ref{th1}, choosing the vector $y=\tfrac{u_\mathcal{V}}{V}-\tfrac{u_{\bar{\mathcal{V}}}}{N-V}$ in \eqref{eq15b} yields a relation between the cut size $\cutsize$ and its inverse-simplex analogue $\cutsizeplus$:
\begin{equation} \label{eq16}
\cutsize \cutsizeplus \geq \left(\frac{V(N-V)}{N}\right)^2.
\end{equation}
Inequality \eqref{eq16} is a generalization of $d_id^{+}_i \geq \left(\tfrac{N-1}{N}\right)^2$ (contained by \eqref{eq16}, for $\mathcal{V}=\lbrace i\rbrace$) which was derived in \cite[Thm. 5]{krl_bestspreader} using algebraic methods rather than geometric ones.
%--------------------------------------
\subsection{Steiner ellipsoid and Laplacian eigenvalues}
\begin{figure}[h!]
\centering
\includegraphics[scale=0.8]{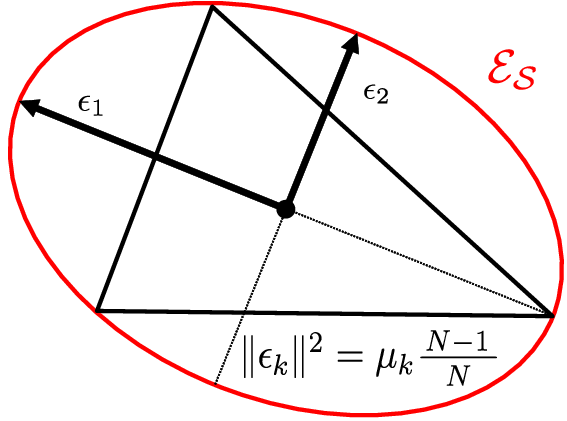}
\caption{The Steiner ellipsoid of a triangle.}
\label{fig10}
\end{figure}

From the eigendecomposition $Q = ZMZ^T$ follows that the Laplacian eigenvectors and eigenvalues contain all information about a graph $G$. Moreover, many important graph properties are captured concisely in terms of the Laplacian eigen-information \cite{pvm_GS}, similar to how some graph properties are easily recognized in the simplex geometry. Interestingly, Fiedler \cite{Fiedler_Steiner} discovered that there is a direct way in which a graph's eigen-information appears in the geometric domain of its corresponding simplex. 
\\
The crucial concept in this correspondence is the \emph{Steiner circumscribed ellipsoid} \cite{Nuesch}. A circumscribed ellipsoid of a simplex $\mathcal{S}$ is an ellipsoid that passes through all vertices of the simplex, and the Steiner circumscribed ellipsoid $\mathcal{E}_\mathcal{S}$ (or simply Steiner ellipsoid) is the unique ellipsoid with minimal volume \cite{Nuesch}. The Steiner ellipsoid is also the unique circumscribed ellipsoid \cite{Fiedler_geometry} of a simplex that has its tangent plane in each of the vertices $s_i$ parallel to the complementary face $\mathcal{F}_{\bar{\lbrace i\rbrace}}$. Given a simplex $\mathcal{S}$, it is thus always possible to find the (unique) Steiner ellipsoid.
\\
The relation between the simplex geometry and the Laplacian eigen-information follows from the fact that for any simplex $\mathcal{S}$, the Steiner ellipsoid $\mathcal{E}_\mathcal{S}$ is given by the points \cite{Nuesch}
\begin{equation} \label{eq18c}
\mathcal{E}_\mathcal{S} = \left\lbrace p\in\mathbb{R}^{N-1} ~\bigg\vert~ p^TS^{\dagger} S^{\dagger T}p = \frac{N-1}{N} \right\rbrace,
\end{equation}
from which Fiedler derived that the semi-axes $\epsilon_k$ of the Steiner ellipsoid are related to the eigenvectors $z_k$ of the Laplacian by \cite[Thm. 6.2.12]{Fiedler_book} (see also Appendix \ref{A Semi-axes}):
\begin{equation}\label{eq19}
\epsilon_k = Sz_k \sqrt{\frac{N-1}{N}}.
\end{equation}
The \emph{semi-axes} of an ellipsoid are the unique set of $N-1$ vectors such that any point $p$ on the ellipsoid can be expressed as $p = \sum_{k=1}^{N-1}\alpha_k\epsilon_k$ with $\sum_{k=1}^{N-1}\alpha_k^2 = 1$. Roughly speaking, the semi-axes diagonalise the ellipsoid (which is a quadric surface). From expression \eqref{eq19} and the fact that $z_k^TQz_k=\mu_k$ we find that the lengths of the semi-axes $\epsilon_k$ of the Steiner ellipsoid $\mathcal{E}_\mathcal{S}$ are proportional to the Laplacian eigenvalues \cite{Fiedler_book}:
\begin{equation} \label{eq19a}
\Vert \epsilon_k \Vert^2 = \mu_k\frac{N-1}{N}
\end{equation}
Figure \ref{fig10} shows an example of a triangle and its corresponding Steiner ellipsoid.
\\
\\
An interesting consequence of relation \eqref{eq19a} between the semi-axes $\epsilon_k$ of the Steiner ellipsoid and the Laplacian eigenvalues $\mu_k$ of the graph follows from the non-uniqueness of the Laplacian eigenvalues. Graphs that share the same eigenvalues but are non-isomorphic are called \emph{cospectral graphs} \cite{VanDam}, and by \eqref{eq19a} their existence also implies that non-congruent simplices can share the same Steiner ellipsoid (which we might call co-Steiner simplices). So, while each simplex $\mathcal{S}$ has a unique Steiner ellipsoid, there might be different simplices that have the same Steiner ellipsoid. Moreover, many classes of cospectral graphs have been identified (for instance most tree graphs), which translates directly to classes of co-Steiner simplices.
\\
\\
\textbf{To summarize:} each simplex $\mathcal{S}$ has a unique circumscribed ellipsoid with minimal volume, called the Steiner ellipsoid $\mathcal{E}_\mathcal{S}$. The (squared) lengths of the semi-axes $\epsilon_k$ of this Steiner ellipsoid are proportional to the Laplacian eigenvalues $\mu_k$ of the graph $G$ corresponding to $\mathcal{S}$, following equations \eqref{eq19} and $\eqref{eq19a}$.
%%%%%%%%%%%%%%%%%
\subsection{Simplex volume and spanning trees}
Another global graph property that appears as a natural geometric feature of the simplex is the number of spanning trees of a graph. A \emph{spanning tree} of a graph $G$ is a connected subgraph of $G$ without cycles. In other words, a spanning tree $T(\widetilde{\mathcal{N}},\widetilde{\mathcal{L}})$ of $G$ is a graph on the same node set as $G$, i.e. $\widetilde{\mathcal{N}} = \mathcal{N}$, and with a link set $\widetilde{\mathcal{L}}\subseteq\mathcal{L}$ such that $T$ is connected and contains no cycles (thus, a tree). In case of a weighted graph, the \emph{weight} of a tree $T$ is equal to the sum of all link weights of $\widetilde{\mathcal{L}}$. Famously, as discovered by Kirchhoff in 1847, \emph{the number of spanning trees $\xi$ of a graph $G$ is proportional to the product of the non-zero Laplacian eigenvalues of $G$} \cite[art. 83]{pvm_GS}:
\begin{equation}\label{eq19b}
\xi = \frac{1}{N}\prod_{k=1}^{N-1}\mu_k.
\end{equation}
For weighted graphs, $\xi$ is defined as the sum of all spanning tree weights and still obeys relation \eqref{eq19b}. Inspired by Fiedler's expression \cite[Cor. 1.4.6]{Fiedler_book} for the volume $\vert\mathcal{S}\vert$ of a simplex $\mathcal{S}$, we derived in \cite{krl_bestspreader} that the volume of $\mathcal{S}$ and of $\mathcal{S}^{+}$ is related to the number of (weighted) spanning trees as
\begin{equation} \label{eq19c}
\vert\mathcal{S}\vert = \frac{N\sqrt{\xi}}{\Gamma(N)} \quad\text{and}\quad \vert\mathcal{S}^{+}\vert = \frac{1}{\Gamma(N)\sqrt{\xi}},
\end{equation}
where $\Gamma(N)$ is the Gamma function. The volume formula \eqref{eq19c} provides interesting insight into qualitative properties of the simplex $\mathcal{S}$ of a graph $G$. For instance, it is known that the complete graph has the most (unweighted) spanning trees of all graphs\footnote{The number of spanning trees of the complete graph  $K_N$ on $N$ nodes is equal to $\xi_{K_N} = N^{N-2}$, a result known as Cayley's formula.}, while a tree graph has only one spanning tree. The relation \eqref{eq19c} between the simplex volume $\vert\mathcal{\mathcal{S}}\vert$ and the number of spanning trees $\xi$ then indicates that the simplex of a complete graph and of a tree are extremal simplices with respect to the volume.
\\
Introducing equation \eqref{eq19a} for the Steiner ellipsoid semi-axis lengths into the formula for the volume of an ellipsoid\footnote{The volume of an ellipsoid $\mathcal{E}$ in $N-1$ dimensions with semi-axis lengths $\alpha_1,\alpha_2,\dots,\alpha_{N-1}$, is equal to $\vert\mathcal{E}\vert = \frac{\pi^{\frac{N-1}{2}}}{\Gamma(\frac{N+1}{2})}\prod_{k=1}^{N-1}\alpha_k$.}, we find that the Steiner ellipsoid volume $\vert\mathcal{E}_\mathcal{S}\vert$ is also related to the number of (weighted) spanning trees as
\begin{equation} \label{eq19d}
\vert\mathcal{E}_\mathcal{S}\vert = \left( \frac{(N-1)\pi}{N}\right)^{\frac{N-1}{2}}\frac{\sqrt{N\xi}}{\Gamma\left(\frac{N}{2} +\frac{1}{2}\right)}.
\end{equation}
Formulas \eqref{eq19c} and \eqref{eq19d} also highlight that the ratio between the volume of a simplex $\vert\mathcal{S}\vert$ and the volume of its circumscribed Steiner ellipsoid $\vert\mathcal{E}_\mathcal{S}\vert$ is independent of the particular simplex, and only depends on the dimension $N-1$ as:
$$
\frac{\vert\mathcal{E}_\mathcal{S}\vert}{\vert\mathcal{S}\vert} = \frac{((N-1)\pi)^{\frac{N-1}{2}}}{N^{\frac{N+1}{2}}}\frac{\Gamma(N)}{\Gamma\left(\frac{N}{2}+\frac{1}{2}\right)} = 
\left(\frac{N-1}{N}\right)^{\tfrac{N}{2}} 
\frac{(2\sqrt{\pi})^N}{2\pi}\frac{\Gamma\left(\tfrac{N}{2}\right)}{\sqrt{N(N-1)}}.
$$
%%%%%%%%%%%%%%%%%%%%%%%%%%%%%%%%%%%

%%%%%%%%%%%%%%%%%%%%%%%%%%%%%%%%%%%
\section{Summary and future directions}\label{S5}
This article gives a self-contained introduction to Fiedler's graph-simplex correspondence \cite{Fiedler_book}, which relates the properties of a weighted, undirected graph $G$ on $N$ nodes, to the geometry of a simplex $\mathcal{S}$ in $\mathbb{R}^{N-1}$. Our description of this correspondence focuses on the role of the Laplacian matrix $Q$ and its pseudoinverse $Q^{\dagger}$ as the key elements connecting simplex geometry to graph theory, and we discuss a number of results that follow from the graph-simplex correspondence:
\begin{itemize}
\item The length of centroid vectors and altitudes in $\mathcal{S}$, as well as inner-products between them, are related to the connectivity structure of $G$: the link weights \eqref{eq11}, degrees \eqref{eq12} and cut sizes \eqref{eq13},\eqref{eq14},\eqref{eq14b} of $G$. Roughly speaking, this connection originates from the fact that these graph properties as well as the simplex properties can be written as a quadratic product of the Laplacian matrix $Q$. As an illustration of the potential use of these results, we connect the Max-Cut problem on graphs to the problem of finding the closest non-intersecting (i.e. complementary) faces in a simplex \eqref{eq14c}.
\item The semi-axes of the Steiner ellipsoid $\mathcal{E}_\mathcal{S}$ of a simplex $\mathcal{S}$ are related to the Laplacian eigenvalues of $G$, as given by \eqref{eq19} and \eqref{eq19a}. This connection is based on the fact that the Steiner ellipsoid is a quadric surface determined by the (positive semidefinite) Laplacian $Q$. As an example, we discuss how equation \eqref{eq19a} relates cospectrality of graphs to the non-bijectivity between a simplex and its Steiner ellipsoid. 
\item The (squared) volume of $\mathcal{S}$ and $\mathcal{E}_\mathcal{S}$ is proportional to the number of spanning trees of $G$, as given by \eqref{eq19c} and \eqref{eq19d}. As a result, simplices with extremal volume can be found from graphs with extremal number of spanning trees, i.e. the complete graph (minimal) and tree graphs (maximal).
\end{itemize}
Finally, since this article presents only a limited account of Fiedler's results, we want to point out three other directions that seem particularly interesting for further investigations:
\begin{itemize}
\item The (squared) distance between two vertices $i$ and $j$ in the inverse simplex is equal \cite[Ch. 6.5]{Fiedler_book} to the effective resistance $\omega_{ij}$, in other words $\omega_{ij} = \Vert s^{+}_i-s^{+}_j\Vert^2$. The \emph{effective resistance} \cite{Klein} is a well-studied graph property related to random walks \cite{Doyle, Chandra}, distances on graphs \cite{Klein_wiener}, network robustness \cite{Ellens} and others, and its direct relation to the geometry of $\mathcal{S}^{+}$ thus seems a promising line of further research.
\item In his early work on graphs and simplices, Fiedler \cite{Fiedler_aggregation} proved an inverse relation\footnote{The inverse relation \cite[Thm. 1.2.4]{Fiedler_book} is defined for block matrices containing $\Omega$ and $Q$, which Fiedler calls the extended Menger matrix and the extended Gram matrix, respectively.} between the effective resistance matrix $\Omega$ of a graph (with elements $(\Omega)_{ij} = \omega_{ij}$), and its Laplacian matrix $Q$. For further details, see for instance \cite[Thm. 1.2.4]{Fiedler_book},\cite{Fiedler_aggregation},\cite{krl_bestspreader}. Fiedler's discovery of the inverse relation between $\Omega$ and $Q$ from 1978 seems to be independent of the derivation by Graham and Lov\'asz \cite{Graham} in 1978 for the inverse of $\Omega$ for a tree, and predates Bapat's formula \cite{Bapat} in 2004 for the inverse of $\Omega$ for general weighted graphs. Moreover, Fiedler's block-matrix inverse formula captures the full structure of the inverse relation between $\Omega$ and $Q$, and we believe that its connection to the geometry of $\mathcal{S}$ can be a valuable tool in the further study of $\Omega$.
\item Fiedler also showed \cite[Thm. 1.3.3]{Fiedler_book} that the angle\footnote{The angle $\theta_{ab}$ between two vectors $a$ and $b$ obeys: $\cos(\theta_{ab}) = \tfrac{a^Tb}{\Vert a\Vert \Vert b\Vert}$. The angle $\phi_{ab}$ between two hyperplanes $\mathcal{H}_a,\mathcal{H}_b$ is equal to $\pi$ minus the angle between the normal vectors $n_a,n_b$ on these hyperplanes, such that $\cos(\phi_{ab}) = -\tfrac{n_a^Tn_b}{\Vert n_a\Vert \Vert n_b\Vert}$ holds.} $\phi^{+}_{ij}$ between two facets $\mathcal{F}^{+}_{\bar{\lbrace i\rbrace}}$ and $\mathcal{F}^{+}_{\bar{\lbrace j\rbrace}}$ in the inverse simplex $\mathcal{S}^{+}$ is related to the graph by $\cos(\phi^{+}_{ij}) = -\tfrac{(Q)_{ij}}{\sqrt{d_id_j}}$. Since angles are natural properties in geometry, this relation might have many interesting implications for graphs and simplices. For instance, non-negativity of the link weights (i.e. $(Q)_{ij}\leq 0$) means that all facet angles in $\mathcal{S}^{+}$ are non-obtuse $\phi^{+}_{ij}\leq \tfrac{\pi}{2}$. Additionally, the relation between the angles $\phi^{+}_{ij}$ and the \emph{normalized Laplacian} $\mathcal{Q}$ of a graph \cite{Chung}, which has elements $(\mathcal{Q})_{ij} \triangleq \tfrac{(Q)_{ij}}{\sqrt{d_id_j}}$, seems interesting to further explore.
\end{itemize}
%%%%%%%%%%%%%%%%%%%%%%%%%%%%%%%%%%%

\bibliographystyle{abbrv}
\bibliography{Bibliography}

%%%%%%%%%%%%%%%%%%%%%%%%%%%%%%%%%%%%%
\appendix
\section*{Appendix}
\section{List of symbols}\label{Alist}
\begin{tabular}{c | l}
\multicolumn{2}{l}{\textbf{Graph-related symbols}} \\ \hline
$G(\mathcal{N},\mathcal{L})$ & Graph with node set $\mathcal{N}$ and link set $\mathcal{L}$ \\ \hline
$N$ & Number of nodes in a graph \\ \hline
$L$ & Number of links in a graph \\ \hline
$w_{ij}$ & Weight of a link between node $i$ and $j$ \\ \hline
$d_i$ & Degree of node $i$ \\ \hline
$Q$ & Laplacian matrix \\ \hline
$\mu_k$ & Laplacian eigenvalue, ordered as $\mu_1\geq\mu_2\geq\dots>\mu_N=0$\\ \hline
$M$ & $(N-1)\times(N-1)$ diagonal matrix containing the non-zero eigenvalues \\ \hline
$z_k$ & Laplacian eigenvector corresponding to $\mu_k$ \\ \hline
$Z$ & $N\times(N-1)$ matrix with the eigenvectors corresponding to non-zero eigenvalues as columns \\ \hline
$\vert\partial\mathcal{V}\vert$ & Cut size, the number of links between nodes in $\mathcal{V}$ and nodes in $\bar{\mathcal{V}}$ \\ \hline
$\xi$ & Number of spanning trees \\ \hline
\multicolumn{2}{l}{\textbf{Simplex-related symbols}} \\ \hline
$\mathcal{S}$ & Simplex \\ \hline
$D$ & Dimension of a simplex; a simplex with $D+1$ vertices is in $\mathbb{R}^D$
\\&In the graph-simplex correspondence, $D=N-1$ \\ \hline
$s_i$ & Vertex vector of vertex $i$ \\ \hline
$S$ & $D\times (D+1)$ matrix with the vertex vectors $s_i$ as columns \\ \hline
$\mathcal{F}_\mathcal{V}$ & Face of the simplex determined by a set $\mathcal{V}$ of vertices\\ \hline
$D_f$ & Dimension of a face; a $D_f$-dimensional face is determined by $V=D_f+1$ vertices.\\ \hline
$x$ & Barycentric coordinate of a point on the simplex; vector in $\mathbb{R}^{D+1}$ with\\&non-negative entries which sum to one \\ \hline
$c_\mathcal{S}$ & Centroid of a simplex; center of gravity \\ \hline
$c_\mathcal{V}$ & Vector pointing to the centroid of $\mathcal{F}_\mathcal{V}$ \\ \hline
$a_\mathcal{V}$ & Altitude; the vector pointing orthogonally between a pair of complementary faces $\mathcal{F}_\mathcal{V}$ and $\mathcal{F}_{\bar{\mathcal{V}}}$ \\ \hline
$\mathcal{E}_\mathcal{S}$ & Steiner circumscribed ellipsoid of a simplex \\ \hline
$\epsilon_k$ & $k^{\text{th}}$ semi-axis of the Steiner ellipsoid \\ \hline
\multicolumn{2}{l}{\textbf{Other symbols}} \\ \hline
$u$  & The all-one vector \\ \hline
$\delta_{ij}$ & The Kronecker delta \\ \hline
$(~)_\mathcal{V}$ & (Vector subscript) Entries not in the set $\mathcal{V}$ are equal to zero, for instance\\&the barycentric coordinate $x_\mathcal{V}$ of points in the face $\mathcal{F}_\mathcal{V}$ \\ \hline
$(~)^{\dagger}$ & (Matrix superscript) Pseudoinverse operator, for instance $Q^\dagger$ and $S^{\dagger}$ \\ \hline
$(~)^{+}$ & (General superscript) Denotes variables related to the pseudoinverse Laplacian and inverse simplex,\\&for instance $s^{+}_i$ and $d^{+}_i = (Q^{\dagger})_{ii}$

\end{tabular}
%%%%%%%%%%%%%%%%%%%%%%%%%%%%%%%%%%%%%%%%%%%%%%%%%%%%%%%%%%%%%%
\section{Halfspace definition of a simplex $\mathcal{S}$} \label{A Halfspace}
Since any point $p$ in $\mathbb{R}^{N-1}$ can be expressed with respect to the simplex vertices as $p = Sy$ for some vector $y\in\mathbb{R}^{N}$, the halfspace inequality \eqref{eq8} can be written as
$$
\mathcal{S} = \left\lbrace p\in\mathbb{R}^{N-1} ~\bigg\vert~ p=Sy\text{~with~}y^TS^TS^{\dagger} \geq -\frac{u}{N}\right\rbrace.
$$
From the pseudoinverse relation \eqref{eq6} between $S$ and $S^\dagger$, and denoting the average value of $y$ by $\bar{y} = \tfrac{u^Ty}{N}$, this yields an elementwise condition on $y$:
\begin{equation}\label{eqA1}
\mathcal{S} = \left\lbrace p\in\mathbb{R}^{N-1} ~\bigg\vert~ p=Sy \text{~with~}(y)_i-\bar{y}+\frac{1}{N}\geq 0 \right\rbrace.
\end{equation}
Since $Su=0$, we can write $p=Sy=S(y-\bar{y}u+\tfrac{1}{N}u)$. The change of variable $y-\bar{y}u+\tfrac{1}{N}u \rightarrow x$ then translates the simplex definition \eqref{eqA1} into the convex hull simplex definition \eqref{eq2}, since $x^Tu=1$ and $(x)_i\geq 0$ hold and $x$ is thus a barycentric coordinate.
\\
A more geometric derivation follows from the fact that each of the facets, e.g. $\mathcal{F}_{\bar{\lbrace i\rbrace}}$, lies on a hyperplane of the form $\lbrace p\in\mathbb{R}^{N-1} \mid p^Ts^{+}_i+\tfrac{1}{N}\geq 0\rbrace\supset \mathcal{F}_{\bar{\lbrace i\rbrace}}$. Each of the elementwise conditions of equation \eqref{eq8}, i.e. $p^Ts^{+}_i \geq -\tfrac{1}{N}$, thus constrains the point $p$ to the inside of one of the $N$ facets $\mathcal{F}_{\bar{\lbrace i \rbrace}}$. The intersection of points that satisfy this condition for all facets is then given by
$$
\mathcal{S} = \bigcap_{i=1}^{N} \left\lbrace p\in\mathbb{R}^{N-1} ~\bigg\vert~ p^Ts^{+}_i \geq -\frac{1}{N} \right\rbrace,
$$
which is equivalent to definition \eqref{eq8}.
%---------------------------------------------
\section{Explicit expression for the altitude $a_\mathcal{V}$} \label{A Altitude}
By definition, the altitude $a_\mathcal{V}$ lies between the complementary faces $\mathcal{F}_\mathcal{V}$ and $\mathcal{F}_{\bar{\mathcal{V}}}$, and is orthogonal to both faces. From the orthogonality property of $a_\mathcal{V}$ and expression \eqref{eq7} for the normals of a face, it follows that the direction $\tfrac{a_\mathcal{V}}{\Vert a_\mathcal{V}\Vert}$ of the altitude must lie in the space $S^{\dagger}x_{\bar{\mathcal{V}}}$ in order to be orthogonal to $\mathcal{F}_{\mathcal{V}}$, and in the space $S^{\dagger}x_{\mathcal{V}}$ in order to be orthogonal to $\mathcal{F}_{\bar{\mathcal{V}}}$, where $x_{\bar{\mathcal{V}}}$ and $x_\mathcal{V}$ are barycentric coordinates for the complementary sets $\bar{\mathcal{V}}$ and $\mathcal{V}$. From these conditions, the following equations follow for the altitude:
\begin{equation}\label{eqA1b}
\frac{a_\mathcal{V}}{\Vert a_\mathcal{V}\Vert} = \frac{S^{\dagger}x^{\star}_{\bar{\mathcal{V}}}}{\sqrt{x^{\star T}_{\bar{\mathcal{V}}} Q^{\dagger} x^{\star}_{\bar{\mathcal{V}}} }} \quad\text{~and~}\quad 
\frac{a_\mathcal{V}}{\Vert a_\mathcal{V}\Vert} = \frac{S^{\dagger}x^{\star}_{\mathcal{V}}}{\sqrt{x^{\star T}_{\mathcal{V}} Q^{\dagger} x^{\star}_{\mathcal{V}}}}.
\end{equation}
Since both equations need to be satisfied simultaneously, we have that the barycentric coordinates $x^{\star}_{\bar{\mathcal{V}}}$ and $x^{\star}_{\mathcal{V}}$ need to determine parallel vectors, i.e. $\tfrac{S^{\dagger}x^{\star}_{\bar{\mathcal{V}}}}{\sqrt{x^{\star T}_{\bar{\mathcal{V}}} Q^{\dagger} x^{\star}_{\bar{\mathcal{V}}} }} = \tfrac{S^{\dagger}x^{\star}_{\mathcal{V}}}{\sqrt{x^{\star T}_{\mathcal{V}} Q^{\dagger} x^{\star}_{\mathcal{V}}}}$ must hold. This condition is only satisfied when $x^{\star}_{\bar{\mathcal{V}}}$ and $x^{\star}_\mathcal{V}$ are equal to the barycentric coordinates of the centroids of the complementary faces $\mathcal{F}_{\bar{\mathcal{V}}}$ and $\mathcal{F}_{\mathcal{V}}$, in other words when $x^{\star}_{\bar{\mathcal{V}}} = \tfrac{u_{\bar{\mathcal{V}}}}{N-V}$ and $x^{\star}_{\mathcal{V}} = \tfrac{u_\mathcal{V}}{V}$. Introducing this solution in equation \eqref{eqA1b} leads to:
$$
\frac{a_\mathcal{V}}{\Vert a_\mathcal{V}\Vert} = \frac{c^{+}_{\bar{\mathcal{V}}}}{\Vert c^{+}_{\bar{\mathcal{V}}}\Vert}. 
$$
Introducing equation \eqref{eq13} for the norm of the inverse centroid then leads to \eqref{eq13c}, and similarly for the altitude $a^{+}_\mathcal{V}$ in the inverse simplex.
%-----------------------------------------------
\section{Proof of Theorem \ref{th1}} \label{A 1-norm inequality}
The $1$-norm of a vector $y\in\mathbb{R}^N$ orthogonal to the all-one vector $u$ can be written as an inner-product:
$$
\Vert y\Vert_1 = \sum_{i=1}^N \vert (y)_i\vert =  (u_{\mathcal{V}_y}-u_{\bar{\mathcal{V}}_y})^Ty,
$$
where $\mathcal{V}_y=\lbrace i\mid (y)_i\geq 0\rbrace$ is the set of non-negative entries of $y$. Since the vector $y$ is orthogonal to $u$, 
we can introduce the matrix $I-\tfrac{uu^T}{N}$ as
$$
\Vert y\Vert_1 = (u_{\mathcal{V}_y}-u_{\bar{\mathcal{V}}_y})^T(I-\tfrac{uu^T}{N})y = 2u_{\mathcal{V}_y}^TS^{\dagger T}Sy,
$$
where the second equality follows from the pseudoinverse relation \eqref{eq6} between $S$ and $S^{\dagger}$, and $u_{\bar{\mathcal{V}}_y} = u-u_{\mathcal{V}_y}$. Invoking the Cauchy-Schwarz inequality \cite[art. 13]{pvm_pA} on this inner-product then yields
$$
\Vert y\Vert_1 \leq 2\sqrt{\left(u_{\mathcal{V}_y}^TQ^\dagger u_{\mathcal{V}_y}\right)\left(y^TQy\right)}.
$$
Since $u_{\mathcal{V}_y}^TQ^\dagger u_{\mathcal{V}_y} = \vert\partial\mathcal{V}_y\vert$, squaring both sides proves Theorem \ref{th1}.\hfill$\square$
%%%%%%%%%%%%%%%%%%%%%%%%%%%%%%%%%%%%%%%%%%%%%
\section{Semi-axes of the Steiner ellipsoid $\mathcal{E}_\mathcal{S}$} \label{A Semi-axes}
We derive expression \eqref{eq19} for the Steiner ellipsoid semi-axes $\epsilon_k$, which shows their relation to the Laplacian eigenvectors $z_k$. The semi-axes of an ellipsoid $\mathcal{E}$ (in $N-1$ dimensions) are the unique set of $N-1$ orthogonal vectors $\epsilon_k$ such that $\mathcal{E}$ can be expressed as
\begin{equation}\label{eqA6b}
\mathcal{E} = \left\lbrace p\in\mathbb{R}^{N-1} ~\bigg\vert~ p = \sum_{k=1}^{N-1} \alpha_k \epsilon_k \text{~with~} \sum_{k=1}^{N-1}\alpha_k^2 = 1 \right\rbrace.
\end{equation}
Starting from equation \eqref{eq18c} for the Steiner ellipsoid, we introduce the transformation $p=Sy$, which translates the condition for $p$ to a condition for $y$ as: $y^T(S^TS^\dagger)(S^{\dagger T}S)y = \tfrac{N-1}{N}$. Since $S^TS^\dagger=I-\tfrac{uu^T}{N}$, the Steiner ellipsoid can be described as
\begin{equation}\label{eqA7}
\mathcal{E}_\mathcal{S} = \left\lbrace p\in\mathbb{R}^{N-1} ~\bigg\vert~ p=Sy \text{~with~}u^Ty=0\text{~and~}\sum_{i=1}^N(y)_i^2 = \frac{N-1}{N} \right\rbrace.
\end{equation}
Next, we consider the projections of $y$ on the $N-1$ Laplacian eigenvectors $z_k$ (excluding $z_N$) as: $y = \sum_{k=1}^{N-1}\beta_kz_k$, where $\beta_k=z_k^Ty$. Since the eigenvectors $z_k$ are orthonormal, $\sum_{k=1}^{N-1}\beta_k^2=\sum_{i=1}^{N}(y)_i^2$ holds for the coefficients $\beta_k$, by which \eqref{eqA7} can be written as
$$
\mathcal{E}_\mathcal{S} = \left\lbrace p\in\mathbb{R}^{N-1} ~\bigg\vert~ p=\sum_{k=1}^{N-1}\beta_k Sz_k \text{~with~} \sum_{k=1}^{N-1}\beta_k^2 = \frac{N-1}{N} \right\rbrace.
$$
Rescaling the coefficients $\beta_k$ by $\sqrt{\tfrac{N}{N-1}}$ and noting that the vectors $Sz_k$ and $Sz_m$ are orthogonal, since $z_k^TQz_m=0$ when $k\neq m$, we find that definition \eqref{eqA7} is equal to the `semi-axes' ellipsoid definition \eqref{eqA6b} when we take $\beta_k\sqrt{\tfrac{N}{N-1}} = \alpha_k$ and $\epsilon_k = Sz_k\sqrt{\tfrac{N-1}{N}}$, proving expression \eqref{eq19} for the semi-axes.
%%%%%%%%%%%%%%%%%%%%%%%%%%%%%%%%%%%%%%%%%%%%%%%%%%%
\end{document}